\documentclass[preprint]{aastex}

\begin{document}


\title{An Indirect Search for Weakly Interacting Massive Particles 
in the Sun Using 3109.6 Days of
  Upward-going Muons in Super-Kamiokande}


\author{
T. Tanaka\altaffilmark{19},
K. Abe\altaffilmark{1,11},
Y. Hayato\altaffilmark{1,11},
T. Iida\altaffilmark{1},
J. Kameda\altaffilmark{1,11},
Y. Koshio\altaffilmark{1,11},\\
Y. Kouzuma\altaffilmark{1},  
M. Miura\altaffilmark{1,11},
S. Moriyama\altaffilmark{1,11},
M. Nakahata\altaffilmark{1,11}
% ...
\\
The Super-Kamiokande Collaboration \\
}
\altaffiltext{1}{Kamioka Observatory, Institute for Cosmic Ray Research, 
University of Tokyo, Kamioka, Gifu, 506-1205, Japan}
% ...
\altaffiltext{11}{Institute for the Physics and Mathematics of the Universe 
(IPMU), The University of Tokyo, Kashiwa, Chiba 277-8568, Japan}
% ...
\altaffiltext{19}{Solar-Terrestrial Environment Laboratory, Nagoya University,
Nagoya, Aichi 464-8602, Japan}
% ...
\footnotetext[1]{$\ddagger$ Deceased}
\footnotetext[2]{$\ast$ Present address: Department of Physics and Astronomy, 
University of Minnesota, MN, 55455, USA}


\begin{abstract}
We present the result of an indirect search for high energy neutrinos from
Weakly Interacting Massive Particle (WIMP) annihilation in the Sun using 
upward-going muon (upmu) events at
Super-Kamiokande. Data sets from SKI-SKIII (3109.6 days) were used for the
analysis. We looked for an excess of neutrino signal from the Sun as compared
with the expected atmospheric neutrino background in three upmu categories:
stopping, non-showering, and showering. No significant excess was observed.
The 90\% C.L. upper limits of upmu flux induced by WIMPs of 100
GeV c$^{-2}$ were 6.4$\times10^{-15}$ cm$^{-2}$ s$^{-1}$ and
4.0$\times10^{-15}$ cm$^{-2}$ s$^{-1}$ for the soft and hard annihilation
channels, respectively.  These limits correspond to upper limits of
4.5$\times10^{-39}$ cm$^{-2}$ and 2.7$\times10^{-40}$ cm$^{-2}$ for
spin-dependent WIMP-nucleon scattering cross sections in the soft and hard
annihilation channels, respectively.
\end{abstract}





\section{Introduction}

From recent observations, Weakly Interacting Massive Particles (WIMPs) are
considered a favorite candidate for cold dark matter ~\citep{wm}. From the
viewpoint of minimum supersymmetric extensions of the Standard Model, the
most well-motivated candidate for WIMPs in the universe is the lightest
supersymmetric neutral particle (LSP) ~\citep{jun, ber}. 
One candidate for the LSP which is theoretically well-studied is the
neutralino ($\widetilde{\chi}$) ~\citep{drees}.  The neutralino is a linear
combination of supersymmetric particles that mix after electroweak-symmetry
breaking.
Taking into account constraints from accelerator experiments and astronomical
observations, the neutralino mass is expected to be in the range from several
tens of GeV to 10 TeV, depending on the model assumed for SUSY breaking and
on the SUSY parameters ~\citep{ams,gil}.  

One method of searching for a WIMP dark matter signal is an indirect search
where decay or annihilation products from WIMPs are observed as originating
from the center of a gravitational potential well such as a celestial
body. WIMPs have a small but finite probability of elastic scattering with a
nucleus. 
As matter is concentrated in deep gravitational potential wells, this
scattering is more likely to occur near the centers of such wells; if their
final velocities after the scattering are less than the escape velocity, they
are trapped gravitationally and eventually accumulated in the center of the
gravitational potential. There are two types of scattering: spin-dependent
(SD) and spin-independent (SI) scattering. In the former, WIMPs couple to the
spin of the target nucleus, and in the latter WIMPs coherently scatter on the
nucleus with probability scaling as the square of the mass of the target
nucleus.  WIMPs, being Majorana particles, can annihilate in pairs, and
produce primarily ${\tau}$ leptons, ${b}$, ${c}$ and ${t}$ quarks, gauge
bosons, and Higgs bosons depending upon their masses and compositions. At the
final step, the decay of the WIMP annihilation products may produce many
kinds of particles such as neutrinos, positrons, antiprotons, antideuterons,
etc.

In the Sun, WIMPs are most likely to scatter on the dominant component,
hydrogen, either via SD or SI coupling. 
Direct detection searches on Earth turn out to have a sensitivity
advantage, generally, for SI interactions, while indirect
searches apply to both~\citep{kami}.
WIMP annihilation products in the form of muon-neutrinos are an
excellent instrument for indirect searches since they can pass through
the matter of the Sun, and interact in the Earth; the resulting detected
muon will be correlated with its parent neutrino direction, providing an
unmistakable signal. In the past, several neutrino telescopes such as
Super-Kamiokande, AMANDA, and IceCube have reported the results of indirect
searches for WIMPs in the Sun~\citep{shan,amanda,ICE}.
Here we investigate upward-going muons (upmus) which are generated
from high energy neutrinos come from the WIMP annihilations 
using the Super-Kamiokande (Super-K, SK).
We search in the direction of the Sun, and an excess of neutrino flux
above the atmospheric neutrino background is sought in the upmu events.
Although recent new detectors such as IceCube have larger acceptances for
higher mass WIMPs, SK is better equipped for
the search of lower mass WIMPs ($<$100 GeV) due to its lower energy threshold
for neutrino signals.

A previous WIMP search in SK was done in 2004 using a dataset of 1679.6 days
of through-going upmu events~\citep{shan}.  In this paper, we have nearly
doubled the size of the dataset (3109.6 days).  In addition, we have divided
the upmu events into three categories, taking into account their neutrino
energy dependence of detector response in order to improve the limit on the
WIMP-induced upmu flux and SD cross section, especially for lower mass WIMPs.
This analysis also now uses WIMPsim~\citep{simm} and the DARKSUSY~\citep{dasu}
simulator in the calculation in order to simulate the WIMP-induced neutrino
fluxes and propagation.

\section{Indirect WIMP search in Super-Kamiokande}

%% In a manner similar to \objectname authors can provide links to dataset
%% hosted at participating data centers via the \dataset{} command.  The
%% second curly bracket argument is printed in the text while the first
%% parentheses argument serves as the valid data set identifier.  Large
%% lists of data set are best provided in a table (see Table 3 for an example).
%% Valid data set identifiers should be obtained from the data center that
%% is currently hosting the data.
%%
%% Note that AASTeX interprets everything between the curly braces in the 
%% macro as regular text, so any special characters, e.g. "#" or "_," must be 
%% preceded by a backslash. Otherwise, you will get a LaTeX error when you 
%% compile your manuscript.  Special characters do not 
%% need to be escaped in the optional, square-bracket argument.
%% In this section, we use  the \subsection command to set off
%% a subsection.  \footnote is used to insert a footnote to the text.

%% This section also includes several of the displayed math environments
%% mentioned in the Author Guide.
The SK detector is a 50-kton water Cherenkov detector located at the Kamioka
mine in Japan. Since SK started operation in 1996, there have been four
experimental phases: SK-I (1996 April--2001 July), SK-II
(2002 October--2005 October), SK-III (2006 July--2008 August), and
SK-IV (2008 September--).  We use the datasets from SK-I to SK-III (3109.6 days)
for this analysis.  

Upmu events, which are categorized as the most energetic events in SK, are
used for the analysis. Muon neutrinos interact in the rock surrounding the
detector to produce muons via charged current interactions. If the energy of
the incoming neutrino is high enough, the muon produced by the neutrino will
enter the detector and emit Cherenkov light.  To separate neutrino-induced
muons from cosmic ray muons, we select only upmus, since the
background from downward-going cosmic ray muons overwhelms any
neutrino-induced muons from above.  Both the neutrino cross section and the
range of the produced muon are proportional to the neutrino energy.  Thus,
the effective target volume is very much increased if we use the upmu events
for high energy neutrino searches.  Upmu events are divided into three
categories: stopping, non-showering through-going, and showering
through-going.
If upmus stop in the detector, they are categorized as
``stopping.'' Stopping upmus are the lowest energy portion of the upmu
sample.  Upmus which fully traverse the detector are classified
as ``through-going.''  The category of through-going is then further
subdivided into ``showering'' (if the event has accompanying radiation due to
radiative energy loss) and ``non-showering'' (otherwise).  The details of the
showering versus non-showering classification scheme are described in
~\citep{shantev}.  Showering through-going events make up the highest
energy events in the upmu sample. The detector angular resolution is
1$^\circ$ for stopping and non-showering through-going, and 1$^\circ$.4 for
showering through-going muons, as estimated by Monte Carlo (MC) simulation. This analysis
is based on a total of 4351 upmus: 919 stopping, 2901
non-showering through-going, and 531 showering through-going. 

The main background source for this analysis is upmu events produced by
atmospheric neutrinos. These were simulated by the NEUT neutrino simulation
package~\citep{neut} and a GEANT-based detector simulator. We use the neutrino
flux calculated by (~\cite{honda} ; $E_{\nu} <$10 TeV) and 
(~\cite{volk} ; $E_{\nu} >$10 TeV) for this simulation.  The main
uncertainty in the absolute number of upmus comes from the uncertainty in the
cosmic ray flux ($\approx20\%$).  In this analysis, this uncertainty is
irrelevant since we normalize the absolute number of simulated atmospheric
neutrino events by the number of upmu data events in each category.  The
effect of neutrino oscillation is also taken into account in the
simulation. We consider $\nu_{\mu} - \nu_{\tau}$ oscillations with $\Delta
m^{2} = 0.0025~\mathrm{eV^{2}}$ and $\sin^{2} \theta = 1.0 $
independently obtained from the previous analyses~\citep{osc1,osc2}.

\section{Results of the WIMP search}
\subsection{Upmu events from the Sun}

% \begin{figure}[!t]
%  \centering
%  \includegraphics[clip,width=3.5in]{cosmod.eps}
%  \caption{Distribution for cos~$\theta_{\rm{sun}}$ of upmu events relative to the
%    Sun. Here, $\cos\theta_{\rm{sun}}=1$ is the direction of the Sun. Crosses
%    indicate the data from SKI-III (3109.6 days) and the histogram indicates
%    atmospheric neutrino MC (taking into account neutrino oscillation)
%    normalized by the total number of data events in each category.}
%  \label{fig:coss}
% \end{figure}

\begin{figure}
  \epsscale{0.7}
  \centering
   \plotone{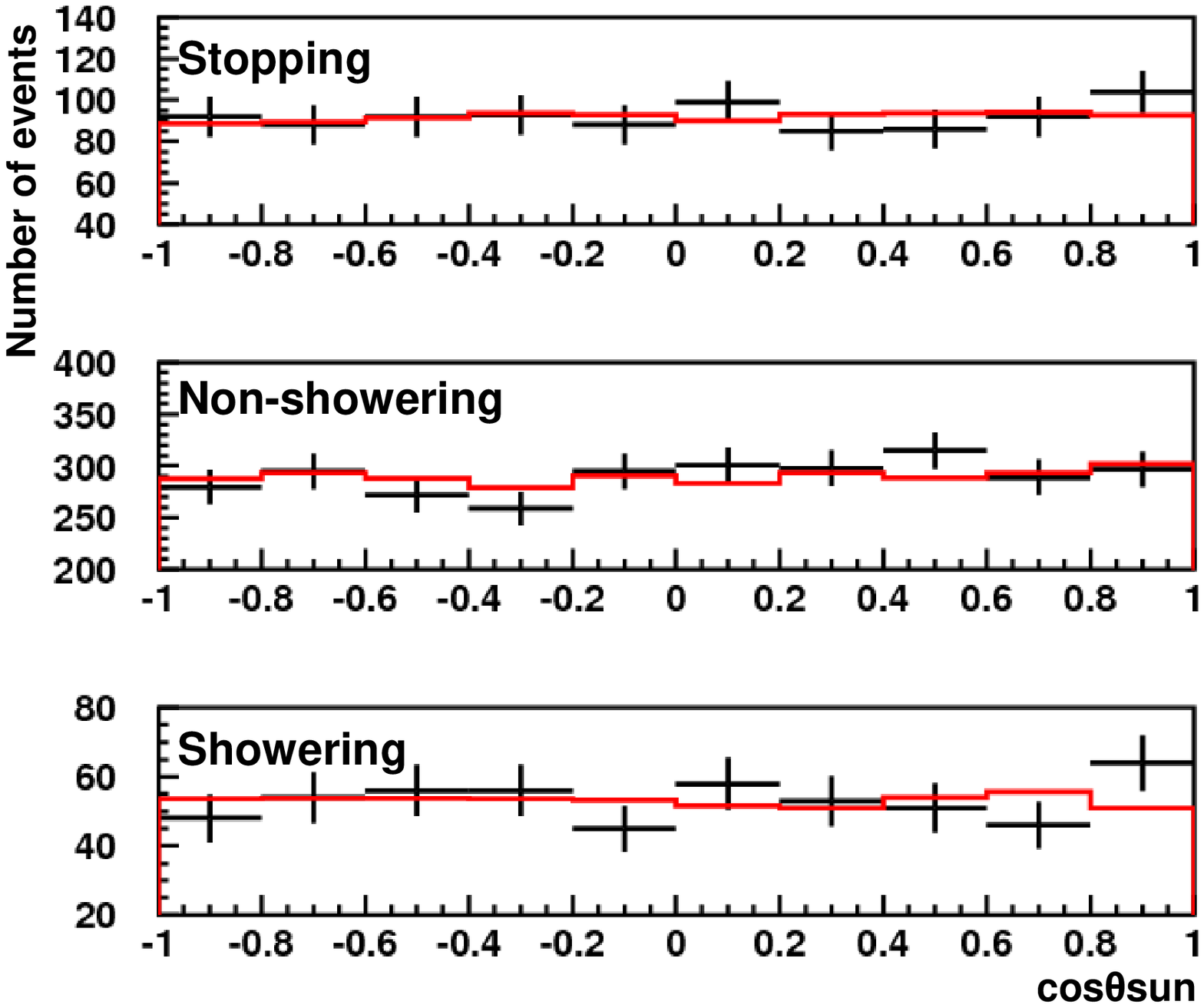}
  \caption{Distribution for cos~$\theta_{\rm{sun}}$ of upmu events relative to the
    Sun. Here, $\cos\theta_{\rm{sun}}=1$ is the direction of the Sun. Crosses
    indicate the data from SKI-III (3109.6 days) and the histogram indicates
    atmospheric neutrino MC with $\nu_{\mu} - \nu_{\tau}$ oscillations 
    normalized by the total number of data events in each category. }
  \label{fig:coss}
 \end{figure}

The event distribution (SKI-III:3109.6 days) as a function of
$\cos\theta_{\rm{sun}}$ for each category of upmu is shown in Figure~\ref{fig:coss}.
Here, $\theta$ is the angle between the reconstructed upmu direction and the
direction of the Sun. This $\theta$ is called the ``cone half-angle.''  In
Figure~\ref{fig:coss}, $\cos\theta_{\rm{sun}}=1$ is defined as the direction of the
Sun. To compare data with the MC events, each MC event was assigned a
time sampled from the observed upmu arrival time in order to match the
livetime distribution and the detector acceptance of the observed
events. The number of MC events is normalized to the total 
number of data events for each category. 
% Y.I. 2011Jun20
Here the normalization was done for the entire $\cos\theta_{\rm{sun}}$ region 
for simplicity.
Without signal regions the difference of normalization is less than 1\%.  

\begin{figure}
   \epsscale{0.5}
  \centering
  \plotone{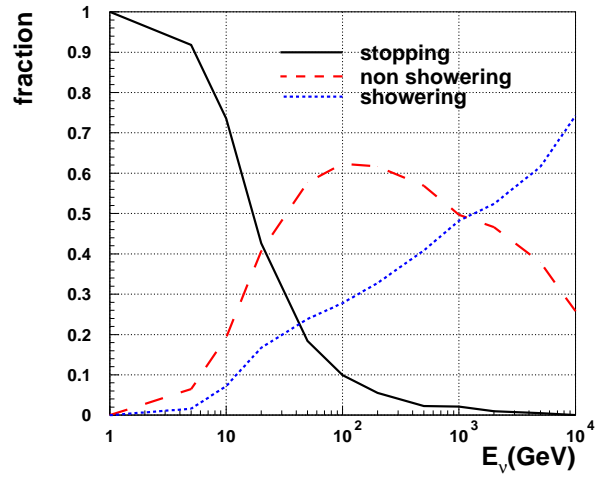}
  \caption{Fraction of each upmu category as a function of the parent
  neutrino energy ($E_{\nu}$). Three upmu categories are shown as follows : 
  stopping (black solid line),
  non-showering (red dashed line), and showering (blue dotted line). The total
 of the fractions for the three categories is normalized to 1. A Monte
  Carlo simulation for
    atmospheric neutrinos is used to calculate this. }
  \label{fig:fraa}
 \end{figure}

We search for an excess from the direction of the Sun in each of the three
upmu categories: stopping (A), non-showering (B), and showering (C).
The fraction of each upmu category as a function of parent neutrino energy
($E_{\nu}$) is shown in Figure~\ref{fig:fraa}. For typical energies of 10, 100,
and 1000~GeV, they are: A:B:C = 0.73\,:\,0.20\,:\,0.07, 0.10\,:\,0.62\,:\,0.28, and
0.02\,:\,0.50\,:\,0.48, respectively. These fractions were obtained from the
atmospheric neutrino MC simulation. The fraction depends only on parent
neutrino energy; its source is irrelevant.  We can roughly deduce the WIMP
mass from the event category where an excess is observed.  Since the spectrum
of the atmospheric neutrino background is proportional to $E^{-2.7}$, while
the signal spectrum from WIMPs gives a bump structure, this energy-dependent
approach is good for reducing the atmospheric neutrino background.
We separately analyze the three upmu samples, then combine the results by
taking into account each fraction assuming a WIMP mass. 

% \begin{figure}[!t]
%  \centering
%  \includegraphics[clip,width=2.8in]{syupap3.eps}
%  \caption{The fraction of each upmu category as a function of the parent
%  neutrino energy ($E_{\nu}$). Three upmu categories are shown: stopping (blac
%k solid line),
%  non-showering (red dashed line), and showering (blue dotted line). The total
% of the
% fractions for the three categories is normalized to 1. A Monte Carlo simulat
%ion for
%    atmospheric neutrinos is used to calculate this. }
%  \label{fig:fraa}
% \end{figure}

\begin{figure}
   \epsscale{0.7}
  \centering
  \plotone{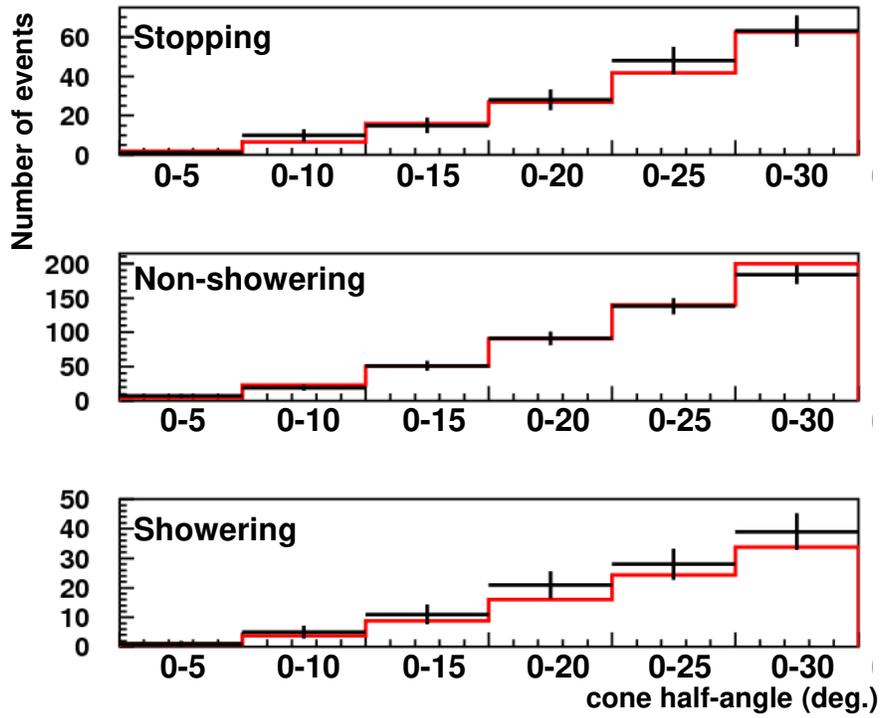}
  \caption{Number of events observed in each cone half-angle toward the
    Sun. Here, 0 deg corresponds to the direction of the Sun. Crosses
    indicate the data from SKI-III (3109.6 days) and the histogram indicates
    atmospheric neutrino MC with $\nu_{\mu} - \nu_{\tau}$ oscillations 
    normalized to the total number of data events in each category.}
  \label{fig:coseha}
 \end{figure}

Since the directional correlation between neutrinos and their produced
leptons depends on neutrino energy, it also depends on the WIMP mass and its
decay mode.  Therefore we look for a possible correlation of upmu direction
to the Sun by searching for an excess of events in various cone half-angles
up to $30^\circ$. Using DARKSUSY, we checked the neutrino-muon angular
correlation for the neutrino spectrum generated by WIMP annihilation in the
Sun for various SUSY parameters and found that the angle between the two is
smaller than $30^\circ$ for most cases.
The number of events observed in each cone half-angle toward the Sun is shown
in Figure~\ref{fig:coseha}. Zero degrees corresponds to the direction of the
Sun. Crosses indicate the data from SKI-III (3109.6 days) and the histogram
indicates atmospheric neutrino MC with same normalization as in
Figure~\ref{fig:coss}. 
%removed by Y.I. 20110715
%Also, we normalized atmospheric neutrino MC excluding the region used
%for the signal search ($\theta_{\rm{sun}}$ $<$ $30^\circ$ ). 
%The normalization factor is varied no more than
% 2 \% at each sample and the chi square value of Fig.~\ref{fig:coseha} varied 
%as 2.65 $\rightarrow$ 3.49 (stoping), 4.01 $\rightarrow$ 2.70
%(non-showering), and 3.12 $\rightarrow$ 4.67 (showering).
%Between two method of normalization, significant difference was not
%found and we judged that no significant excess is observed. 
%To act more conservative analysis, the former normalization which uses
%all theta region of dataset is used for following analyses. 

% \begin{figure}[!t]
%  \centering
%  \includegraphics[clip,width=3.5in]{conehalf.eps}
%  \caption{Number of events observed in each cone half-angle toward the
%    Sun. Here, 0 degrees corresponds to the direction of the Sun. Crosses
%    indicate the data from SKI-III (3109.6 days) and the histogram indicates
%    atmospheric neutrino MC (taking into account neutrino oscillation)
%    normalized to the total number of data events in each category.}
%  \label{fig:coseha}
% \end{figure}

\subsection{Upmu flux limit from the Sun }
\begin{figure}
  \epsscale{0.6}
  \centering
  \plotone{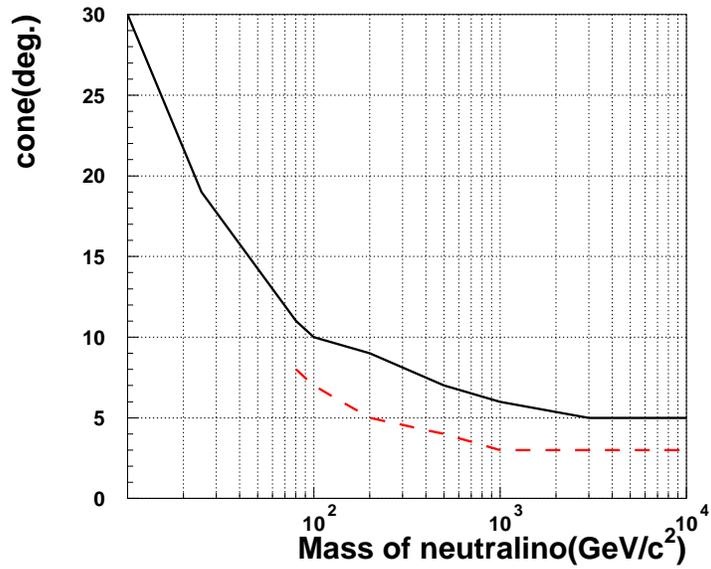}
  \caption{Cone half-angle which contains more than 90\% of the total number
    of upmus expected from WIMP annihilations in the Sun when we assume
 the soft annihilation channel (black solid line) and hard annihilation
 channel (red dashed line). This result was obtained from WIMPsim.}
  \label{fig:shcone}
 \end{figure}

We calculate the upper limit of upmu flux from the direction of the Sun with
respect to the WIMP mass.  Here we estimate the cone half-angles which
contain more than 90\% of the total number of upmus expected from WIMP
annihilations in the Sun.
We use the WIMPsim package~\citep{simm}, which simulates WIMP annihilations,
propagation of neutrinos from those annihilations, and the propagation of
muons after neutrino interactions in the rock.  In order to derive the
expected neutrino spectrum from the WIMP annihilation, we consider two cases:
annihilation into $b\bar{b}$~(soft channel), and annihilation into
$W^{+}W^{-}$~(hard channel).
These give the softest and hardest neutrino spectra, respectively.  The cone
half-angles obtained from these spectra are shown in
Figure~\ref{fig:shcone}. The cone half-angle for the soft channel is generally
wider at the same WIMP mass since many neutrinos from annihilation via the
soft channel are expected to have lower energy.

%YI 2011Jun20
The other smearing effects for angular correlation are muon propagation 
in the surrounding rocks, distribution of WIMPs in the Sun, and 
angular resolution of upmu reconstruction. 
The smearing due to Coulomb multiple scattering in the surrounding rocks 
is 2$^\circ$ -- 3$^\circ$ for muons with energy above 1 GeV.
The angular size of the Sun at the Earth is 0$^\circ$.5, which is 
negligible~\citep{joamas}. 
These two smearing effects were taken into account by WIMPsim.
The detector angular resolution of 1$^\circ$ (stopping and non-showering) 
and 1$^\circ$.4 (showering) are smaller than the smallest cone half angle 
and thus were not taken into account.
%YI

%remvoed by Y.I. Jul11,2011
%The number of expected upmu events in each cone half-angle is affected by
%uncertainties such as: muon propagation in the surrounding rocks,
%distribution of WIMPs in the Sun, and angular resolution of upmus.  As for
%the muon propagation, the angular resolution of the muon direction due to
%multiple Coulomb scattering is $\approx 2^\circ - 3^\circ$ for muon energies
%above 1 GeV. Since the angular size of the Sun at Earth is 0.5$^\circ$, we
%take the Sun as a point-like source~\citep{joamas}.
%These effects are also taken into account in WIMPsim and reflected in
%Fig.~\ref{fig:shcone}.
%The reconstruction uncertainty is not taken into account. Even at the 
%smallest cone half angle used for this analysis (3$^\circ$ at 
%m$_\chi$=10000 GeV/c$^2$, hard channel), the 
%reconstruction uncertainty broaden it by $\sim$ 0.5$^\circ$. 
%Thus, we ignore the reconstruction uncertainty.
%removed by Y.I.

We show the cone half-angle for several WIMP masses in
 Tables~\ref{tab:flim_sof}
and \ref{tab:flim_har}.  Table~\ref{tab:flim_har} begins at 80.3 GeV since
WIMPs with mass lower than 80.3 GeV cannot annihilate into $W^{+}W^{-}$.
We followed the DarkSUSY for the treatment of masses of standard model 
particles ~\citep{simm}.

\begin{figure}
  \epsscale{1.0}
  \plottwo{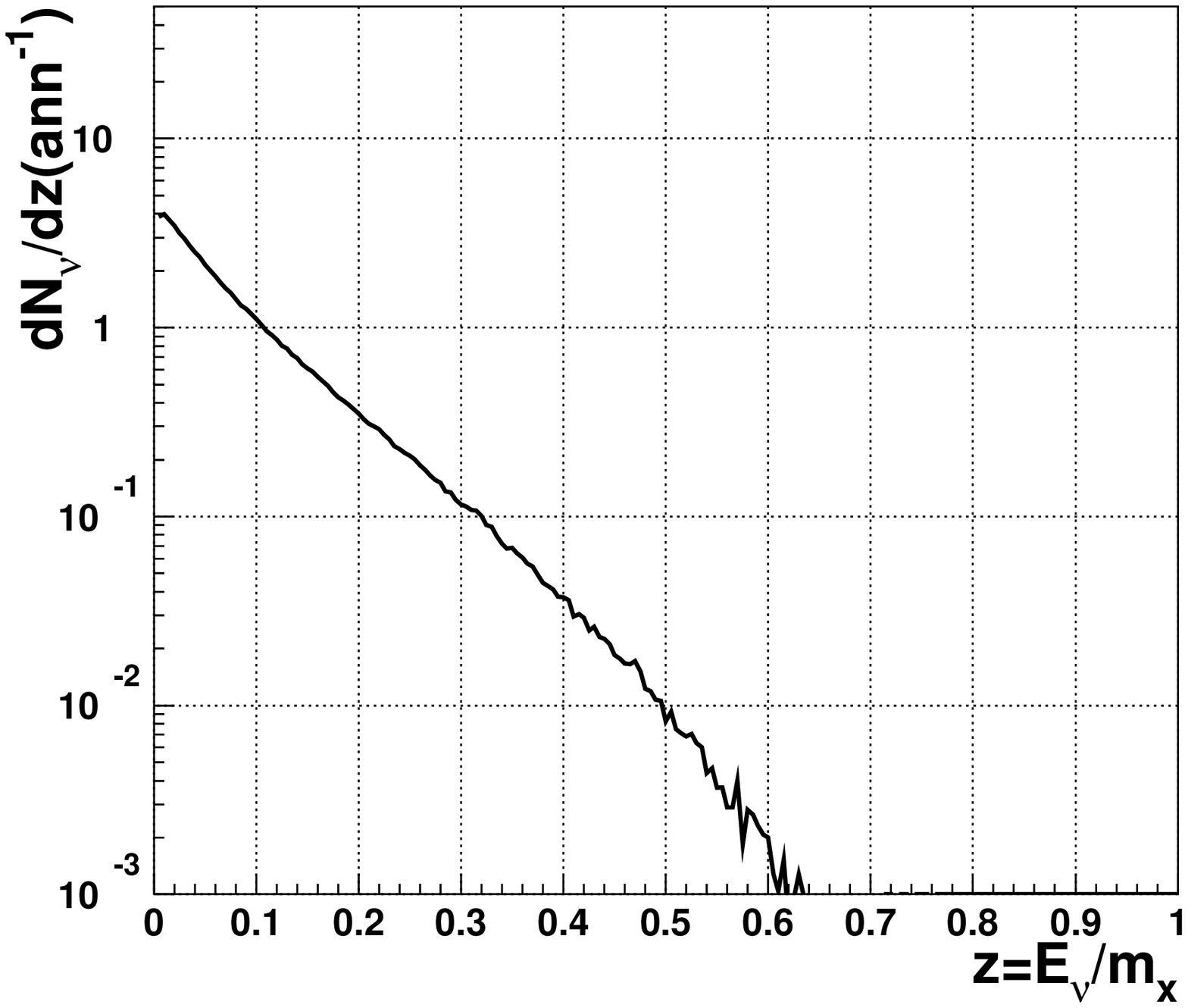}{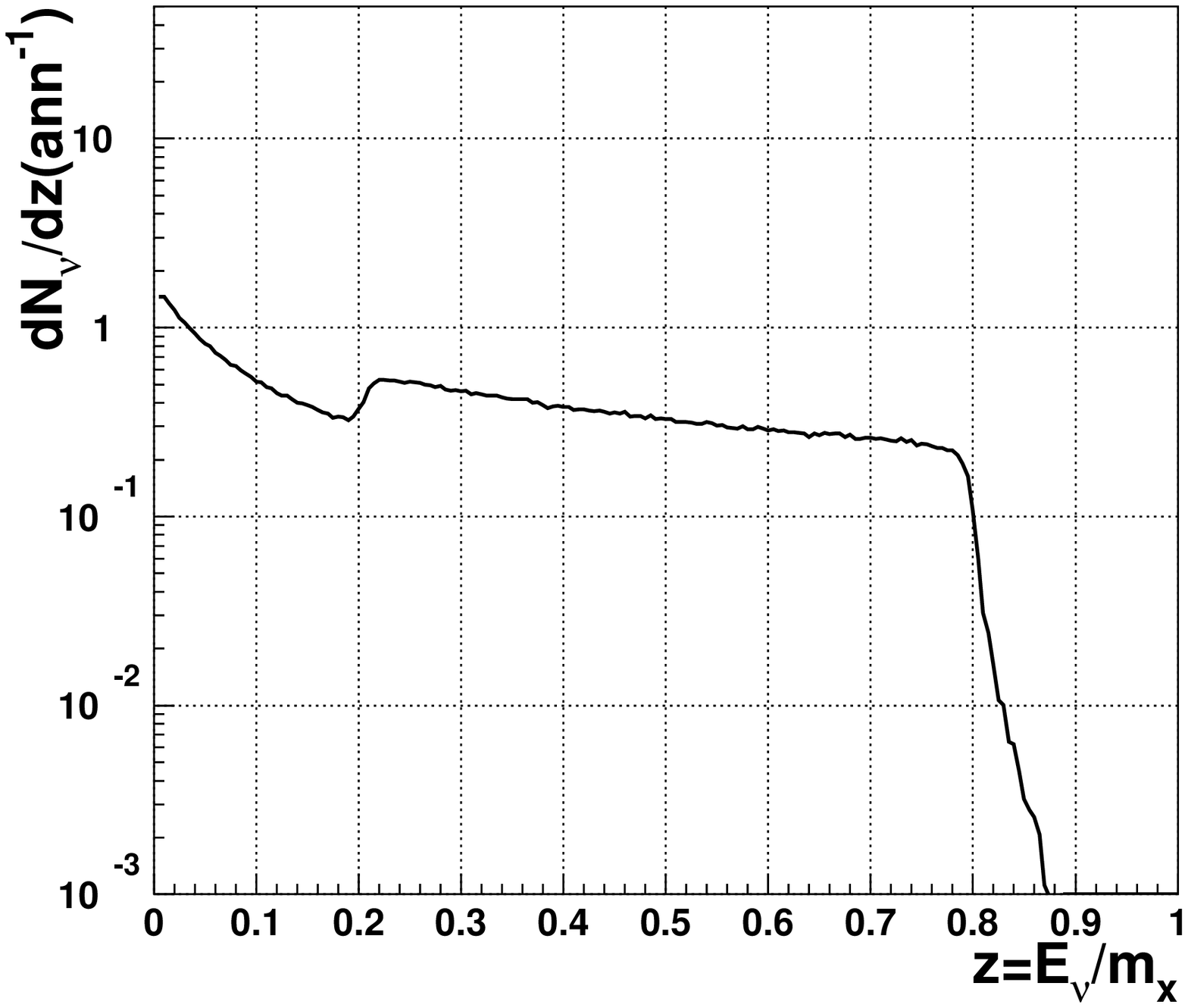}
  \caption{Simulated neutrino flux ($\nu_{\mu} + \bar{\nu_{\mu}}$) at the
  surface of the detector from the annihilation of neutralinos of 100
  GeV in the soft annihilation channel alone (left panel) and 
  hard annihilation channel alone (right panel).}
  \label{fig:sofspec}
 \end{figure}

For each cone half-angle, which corresponds to a certain WIMP mass in each
annihilation channel, we calculated the upper Poissonian limit of the
upmu flux:

%Y.I. Jul11,2011
\begin{equation}
\phi(90\% \rm{C.L.}) = \frac{N_{90}(n^1_{\rm{obs}} ,n^1_{\rm{BG}}, F^1, 
n^2_{\rm{obs}} ,n^2_{\rm{BG}}, F^2, n^3_{\rm{obs}} ,n^3_{\rm{BG}}, F^3)}
{\epsilon \times A \times T},   
\label{neutlino}
\end{equation}
%

%removed by Y.I.
%\begin{equation}
%\phi(90\% C.L.) = \frac{N_{90}(n^{1}_{obs} ,n^{1}_{BG}, F^{1}, 
%n^{2}_{obs} ,n^{2}_{BG}, F^{2},n^{3}_{obs},
%n^{3}_{BG}, F^{3} )}
%{\epsilon \times A \times T},   
%\label{neutlino}
%\end{equation}
%

\noindent 
where $A$ is the detector cross-sectional area in the direction of the
expected signal averaged over Sun directions. For each position of the Sun,
this area is calculated using a minimal track length of 7m in the inner
detector as a criterion to select viable upmu trajectories. It varies from
$968~\mathrm{m^{2}}$ to $1290~\mathrm{m^{2}}$ with the position of the
Sun. $T$ is the experimental livetime and $\epsilon$ is the detection
efficiency. For upmu samples, the inefficiency of detection is
negligible~\citep{shantev}. The average exposure $A \times T$ is 
$1.67 \times 10^{11}~\mathrm{m^{2}s}$. 
$N_{90}$ is the upper Poissonian limit on
the number of upmu events at 90\% C.L., given the number of observed events
of each category ($n^i_{\rm{obs}}$), 
expected background of each category
($n^i_{\rm{BG}}$) 
and fraction of each category
($F^i$) where index $i$ corresponds three types of upmu category.  
This fraction is estimated from the neutrino spectrum of the
given WIMP mass for each annihilation channel and the fraction of the upmu
categories as a function of the parent neutrino energy as shown in
Figure~\ref{fig:fraa}.  The neutrino spectra were simulated by WIMPsim.
Figure~\ref{fig:sofspec} shows the simulated neutrino
spectra for the soft and hard annihilation channels, assuming a neutralino
mass of 100 GeV.

% \begin{figure}[!t]
%  \centering
%  \includegraphics[clip, width=2.5 in]{100gev_soft.eps}
%  \caption{Simulated neutrino flux ($\nu_{\mu} + \bar{\nu_{\mu}}$) at the
%  surface of the detector from the annihilation of neutralinos of 100
%  GeV in the soft annihilation channel alone.}
%  \label{fig:sofspec}
% \end{figure}

% \begin{figure}[!t]
%  \centering
% \includegraphics[clip,width=2.5 in]{100gev_hard.eps}
% \caption{Simulated neutrino flux ($\nu_{\mu} + \bar{\nu_{\mu}}$) at the
%  surface of the detector from the annihilation of neutralinos of 100
%  GeV in the hard annihilation channel alone.}
%  \label{fig:harspec}
% \end{figure}

 $N_{90}$ was calculated combining all categories of upmu for a certain mass
 based on Bayes' theorem~\citep{kari}.  The Poisson-based likelihood function
 for the expected signal is as follows :

%Y.I. Jul11, 2011
\begin{equation}
L(n^i_{\rm{obs}} | \nu_s) = \prod_{i=0}^{3}  \frac{(\nu_s F^i + n^i_{\rm{BG}})^{n^i_{\rm{obs}}}}{n^i_{\rm{obs}} ! } e^{-(\nu_s F^i + n^i_{\rm{BG}})} \label{n90eq1},
\end{equation}
%

%removed by Y.I. Jul11,2011 
%\begin{equation}
%L(n^i_{obs} | \nu_s) = \prod_{i=0}^{3}  
%\frac{(\nu_s F^i + n^i_{BG})^{n^i_{obs}}}{n^i_{obs} ! }
% e^{-(\nu_s F^i + n^i_{BG})} \label{n90eq1}
%\end{equation}
%

\noindent
where $\nu_s$ is the expected real signal.
Using the likelihood function in Equation (~\ref{n90eq1}), the upper
confidence level is written as

 \begin{equation}
\rm{C.L.} = \frac{ \int_{\nu_{s}=0}^{N_{\rm{limit}}}L(n^i_{\rm{obs}} | \nu_s) d\nu_{s}}
{\int_{\nu_{s}=0}^{\infty} L(n^i_{\rm{obs}} | \nu_s) d\nu_{s}}. \label{n90eq2}
\end{equation}

\noindent
Applying 90\% as the C.L. in Equation (~\ref{n90eq2}), the upper Poissonian
limit ($N_{90}$) is obtained.

%%%Table

The summaries of the numbers of upmu events and the estimated backgrounds for
the cone half-angles which correspond to each WIMP mass, $N_{90}$, and the
calculated flux limits are shown in Tables \ref{tab:flim_sof} and \ref{tab:flim_har}.

\begin{figure}
   \epsscale{0.6}
  \centering
  \plotone{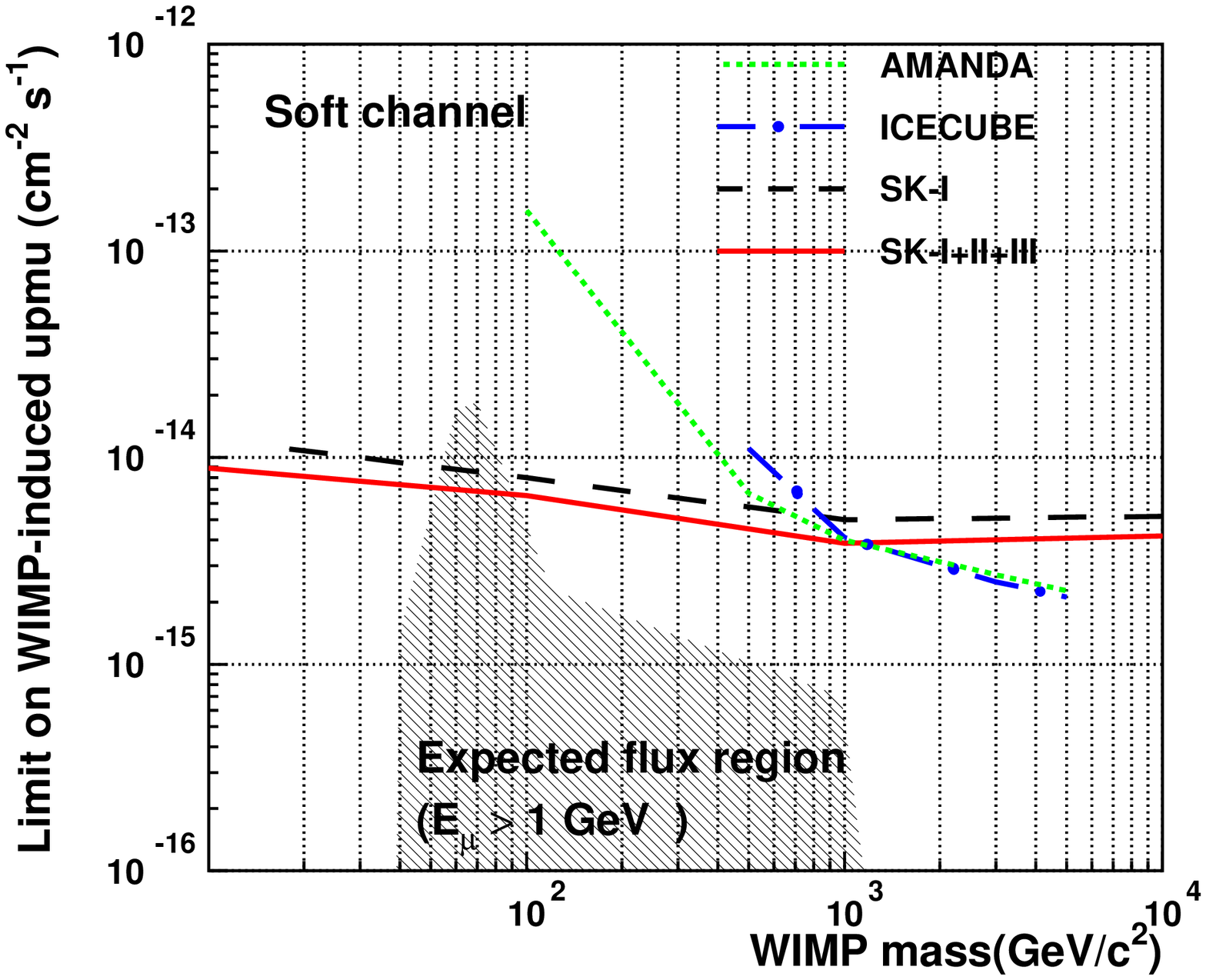}
  \caption{Upmu flux limit vs. WIMP mass for the soft annihilation
    channel. The flux limits of AMANDA (~\cite{amanda}, ; green dotted line ),
    IceCube (~\cite{ICE},; blue dot-dash line), and this analysis (red solid
    line) are shown.  Also shown is the previous limit from
    Super-K (~\cite{shan}, ; black dashed line).  Note that SK's previous limit
    does not assume the pure soft channel, but assumes a typical annihilation
    channel in which the branching ratio for the soft annihilation channel is
    relatively high (80\%). The shaded region is the expected flux region
    from DARKSUSY.}
  \label{fig:flim_sof}
 \end{figure}

\begin{figure}
  \centering
  \plotone{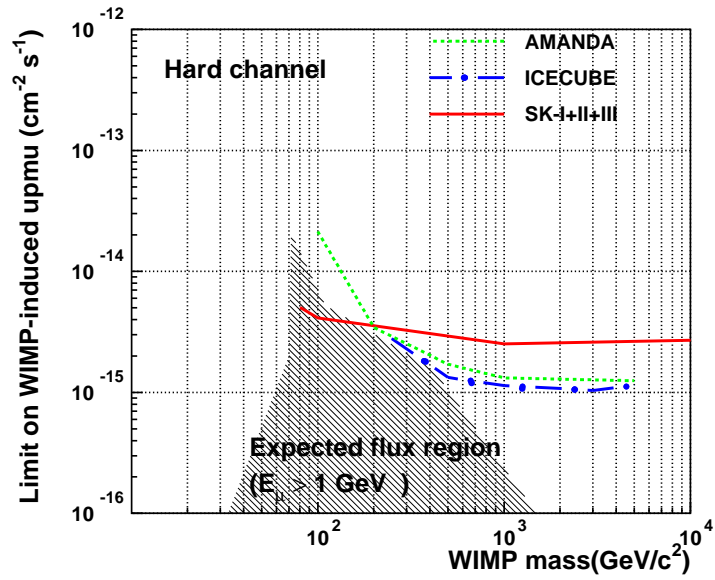}
  \caption{Upmu flux limit vs. WIMP mass for the hard annihilation
    channel. The flux limits of AMANDA (~\cite{amanda}, ; green dotted line),
    IceCube (~\cite{ICE}, ; blue dot-dash line), and this analysis (red solid
    line) are shown. The shaded region is the expected flux region from
    DARKSUSY.}
  \label{fig:flim_har}
 \end{figure}

The 90\% confidence level flux limits as a function of WIMP mass are shown in
Figures~\ref{fig:flim_sof} and \ref{fig:flim_har}.  Results from other
experiments, AMANDA~\citep{amanda} and IceCube ~\citep{ICE}, are also shown in
these figures.  The previous Super-K result~\citep{shan} is also shown in
Figure~\ref{fig:flim_sof}, but please note that it was calculated assuming an
annihilation branching ratio.  That result assumed that 80\% of the
annihilation products were $b\bar{b}$, 10\% were $c\bar{c}$, and 10\% are
$\tau\bar{\tau}$, as a typical branching ratio. We cannot regard this as a
pure soft annihilation, but it is shown for reference.

% \begin{figure}[!t]
%  \centering
%  \includegraphics[clip,width=3.5 in]{soft_flimit.eps}
%  \caption{Upmu flux limit vs. WIMP mass for the soft annihilation
%    channel. The flux limits of AMANDA~\cite{amanda} (green dotted line),
%    IceCube~\cite{ICE} (blue dot-dash line), and this analysis (red solid
%    line) are shown.  Also shown is the previous limit from
%    Super-K~\cite{shan} (black dashed line).  Note that SK's previous limit
%    doesn't assume the pure soft channel, but assumes a typical annihilation
%    channel in which the branching ratio for the soft annihilation channel is
%    relatively high (80\%). The shaded region is the expected flux region
%    from DARKSUSY.}
%  \label{fig:flim_sof}
% \end{figure}

% \begin{figure}[!t]
%  \centering
%  \includegraphics[clip,width=3.5 in]{hard_flimit.eps}
%  \caption{Upmu flux limit vs. WIMP mass for the hard annihilation
%    channel. The flux limits of AMANDA~\cite{amanda} (green dotted line),
%    IceCube~\cite{ICE} (blue dot-dash line), and this analysis (red solid
%    line) are shown. The shaded region is the expected flux region from
%    DARKSUSY.}
%  \label{fig:flim_har}
% \end{figure}

\subsection{Limit on WIMP-proton SD cross section}
From the upmu flux limits shown in Figures~\ref{fig:flim_sof} and
\ref{fig:flim_har}, we also calculate the upper limit on the
SD scattering cross section of WIMPs.  If we assume that the
capture rate of WIMPs and the WIMP annihilation rate are in equilibrium in
the Sun, the neutrino flux from annihilation depends only on the capture
rate.  Since the age of the Sun is much longer than the timescale of the
annihilation, it is a safe assumption that they are in equilibrium.  We can
relate the  upmu flux, $\Phi_{\mu}^{f}$, to the SD cross 
section, $\sigma^{\rm{SD}}$, on the assumption that only one annihilation channel
is dominant and the SI cross section is zero.  The SD cross section,
$\sigma^{\rm{SD}}$, can be written as~\citep{gus}

\begin{equation}
\sigma^{\rm{SD}} = \kappa_{f}^{\rm{SD}}(m_{\chi})\Phi_{\mu}^{f} ,  \label{relete}
\end{equation}

\noindent
where $\kappa_{f}^{\rm{SD}}(m_{\chi})$ is the conversion factor between the SD cross
section and the muon flux. This value depends on the WIMP mass, $m_{\chi}$.
We use the factor introduced in Figure 3 of \citep{gus}, which was
calculated using DARKSUSY.
% Removed by Y.I. 20110715
%The main uncertainties which affect the limit on WIMP-proton
%SD cross section are come from annihilation branching ratio and 
%astrophysical uncertainties~\citep{gus}.
% They are considered to be bigger than 
%overall detector uncertainties which is considered to be $\sim$ 10 \% 
%~\citep{roge} and we 
%ignore the detector uncertainties in this calculation.

% \begin{figure*}[tb]
%  \centering
%  \includegraphics[clip,width=4.5 in]{sdlimitnov.eps}
%  \caption{Limit on the WIMP-proton spin-dependent cross section as a
%    function of WIMP mass.  Limits from direct detection experiments:
%    DAMA/LIBRA allowed region~\cite{dama} (dark red and light red filled, for
%    with and without ion channeling, respectively), KIMS~\cite{kims} (light
%    blue crosses), and PICASSO~\cite{pica} (grey dotted line) are shown.
%    Also we show here the results of indirect detection (neutrino
%    telescopes): AMANDA~\cite{amanda} (black line with triangles),
%    IceCube~\cite{gus} (blue line with squares), and this analysis (red line
%    with stars). Also the previous limit from Super-K (green dashed line) is
%    shown. This plot was made from~\cite{tool}.}
%  \label{fig:cross}
% \end{figure*}

\begin{figure}
  \centering
  \plotone{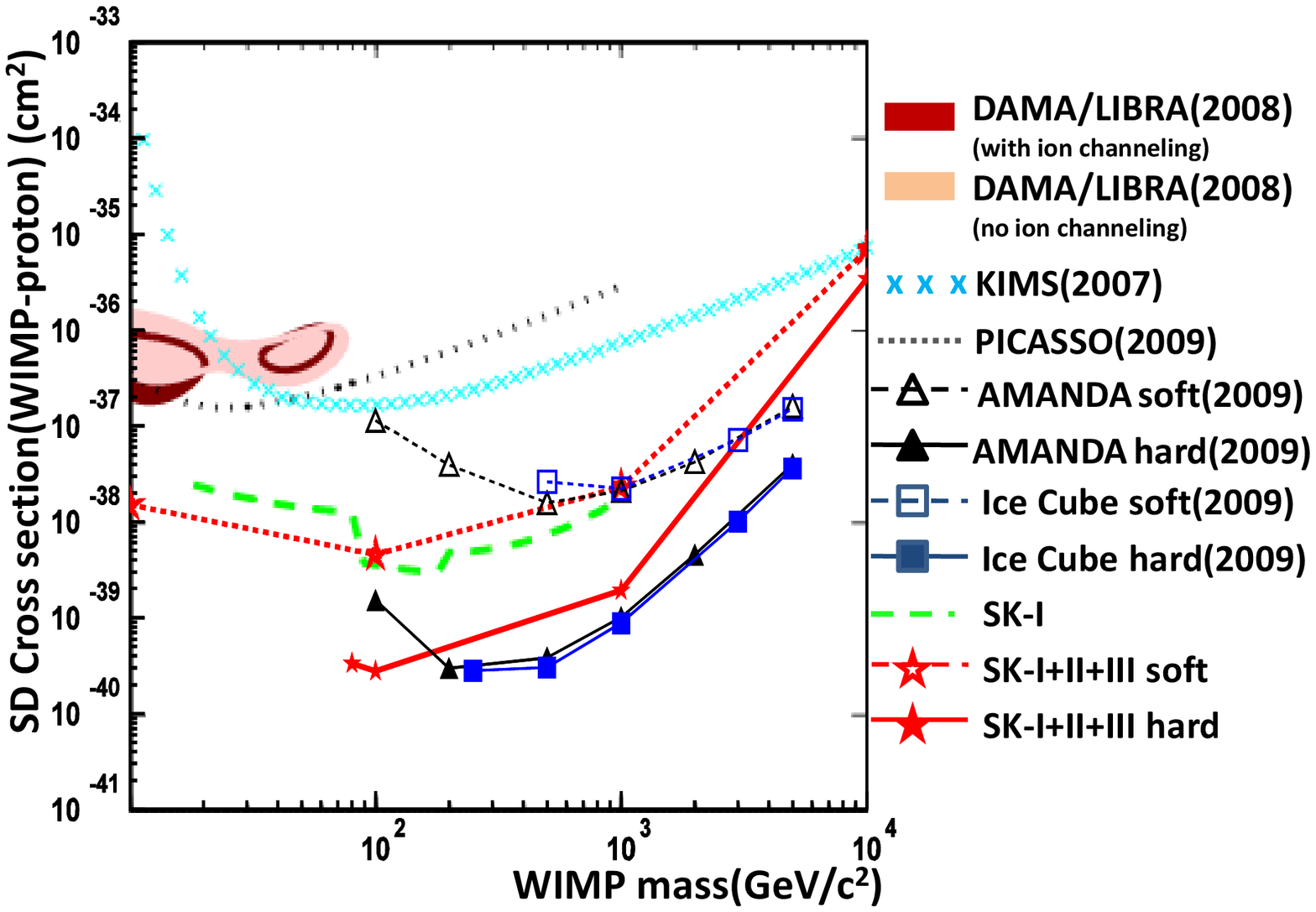}
  \caption{Limit on the WIMP-proton spin-dependent cross section as a
    function of WIMP mass.  Limits from direct detection experiments:
    DAMA/LIBRA allowed region (~\cite{dama}, ; dark red and light red filled, for
    with and without ion channeling, respectively), KIMS (~\cite{kims}, ; light 
    blue crosses), and PICASSO (~\cite{pica}, ; grey dotted line) are shown.
    Also we show here the results of indirect detection (neutrino
    telescopes): AMANDA (~\cite{amanda}, ; black line with triangles),
    IceCube (~\cite{gus}, ; blue line with squares), and this analysis (red line
    with stars). Also the previous limit from Super-K (green dashed line) is
  shown. This plot was made from SUSY Dark Matter/Interactive Direct
  Detection Plotter, http://dmtools.berkeley.edu/limitplots .}
  \label{fig:cross}
 \end{figure}

The calculated upper limit is shown in Figure~\ref{fig:cross}.  Also shown are
results from direct detection experiments: DAMA/LIBRA allowed
region~\citep{dama}, KIMS~\citep{kims}, and PICASSO~\citep{pica}; and from
neutrino telescopes: AMANDA~\citep{amanda} and IceCube~\citep{ICE}.  All of the
limits from the neutrino telescopes (including our limit) are calculated by
the same method described in \citep{gus}. The previous limit from
Super-K is also shown for reference.  Note that the gaps at WIMP masses of
$\sim$ 80~GeV and 174~GeV in this previous result are due to the thresholds
of the $W^{+}W^{-}$ and $\tau\bar{\tau}$ annihilation channels. These do not
appear in the new result because we assume only one annihilation channel.
Our limit is improved at low WIMP mass (below 100 GeV), which is stronger
than limits from other indirect-detection experiments. Our limit is also
stronger than direct-detection experiments.

\section{Conclusion}
We have performed an updated indirect search for WIMPs using the
Super-Kamiokande data of SKI-III (3109.6 days).  A WIMP-induced neutrino
signal was searched for in the direction of the Sun.

We compared upmu events with the estimated background of
atmospheric neutrinos. No significant excess was observed. 

To calculate the upmu flux limit, we used a new method involving WIMPsim and
the DARKSUSY simulator. Cone half-angles which contained more than 90\% upmu
signal for each WIMP mass were extracted from simulation.  Upper upmu flux
limits as a function of WIMP mass for the soft and the hard annihilation
channels were obtained.  Our limit extends into the region of expected upmu
flux estimated by some SUSY models.

An upper limit on the SD scattering cross section was also
obtained. We used a WIMP mass-dependent conversion factor as described
in \citep{gus}.
%We got the better limit comparing with other experiment of direct detection
%and at the range of low mass WIMPs, lower limit was obtained comparing
%with the result from the neutrino telescopes. 
At low WIMP mass (below 100~GeV) our limit on the WIMP-proton
SD cross section is better than other experiments for both
direct-detection and indirect-detection experiments.

\clearpage

%% the file, uncomment the \plotone, \plottwo, and
%% If you need a layout that cannot be achieved with \plotone or
%% \plottwo, you can invoke the graphicx package directly with the
%% \includegraphics command or use \plotfiddle. For more information,
%% please see the tutorial on "Using Electronic Art with AASTeX" in the
%% documentation section at the AASTeX Web site,
%% http://www.journals.uchicago.edu/AAS/AASTeX.
%%
%% The examples below also include sample markup for submission of
%% supplemental electronic materials. As always, be sure to check
%% the instructions to authors for the journal you are submitting to
%% for specific submissions guidelines as they vary from

%% This example uses \plotone to include an EPS file scaled to
%% 80% of its natural size with \epsscale. Its caption
%% has been written to indicate that additional figure parts will be
%% available in the electronic journal.

%\clearpage

%% Here we use \plottwo to present two versions of the same figure,
%% one in black and white for print the other in RGB color
%% for online presentation. Note that the caption indicates
%% that a color version of the figure will be available online.
%%
%% This figure uses \includegraphics to scale and rotate the still frame
%% If you are not including electonic art with your submission, you may
%% mark up your captions using the \figcaption command. See the
%% No more than seven \figcaption commands are allowed per page,
%% so if you have more than seven captions, insert a \clearpage

%% Tables should be submitted one per page, so put a \clearpage before

%% Two options are available to the author for producing tables:  the
%% deluxetable environment provided by the AASTeX package or the LaTeX
%% table environment.  Use of deluxetable is preferred.
%%

%% Three table samples follow, two marked up in the deluxetable environment,
%% one marked up as a LaTeX table.

%% In this first example, note that the \tabletypesize{}
%% command has been used to reduce the font size of the table.
%% We also use the \rotate command to rotate the table to
%% landscape orientation since it is very wide even at the
%% Note also that the \label command needs to be placed

%% This table also includes a table comment indicating that the full
%% version will be available in machine-readable format in the electronic

\clearpage

%% If you use the table environment, please indicate horizontal rules using
%% Do not put multiple tabular environments within a single table.
%% The optional \label should appear inside the \caption command.

\begin{table}
\begin{center}
\caption{Event summary to calculate the flux limit ($\phi$) assuming the
   soft annihilation channel is dominant.  The numbers of events ($n_{\rm{obs}}$)
   and estimated background ($n_{\rm{BG}}$) for each cone half-angle ($\theta$)
   which corresponds to a certain WIMP mass ($m_\chi$) are shown for all
   categories of upmu. This cone half-angle is defined by the criterion that
   more than 90\% of the WIMP-induced upmus are contained for the assumed
   WIMP masses.  Also we show the estimated fraction ($F$) of each upmu
   category for the given WIMP mass.  The calculated result of $N_{90}$ and the
   flux limit for each WIMP mass are shown in the last two columns. 
 \label{tab:flim_sof}}
\begin{tabular}{ccccccccccccc}
\tableline\tableline
  {} & {} &\multicolumn{3}{c}{Stopping} &
      \multicolumn{3}{c}{Non-showering}  &
      \multicolumn{3}{c}{Showering} &\multicolumn{2}{c}{} \\ \tableline

 {$m_\chi$} & {$\theta$} &{$n_{\rm{obs}}$} &{$n_{\rm{BG}}$}  &
  {$F$} &{$n_{\rm{obs}}$} &{$n_{\rm{BG}}$} &{$F$} &
  {$n_{\rm{obs}}$} &{$n_{\rm{BG}}$}  &{$F$} &{$N_{90}$}  &
  {$\phi (\times10^{-15})$}
    
           \\
  {(GeV)} & {(deg)} &{} &{}  &
    {($\%$)} &{} &{} &{($\%$)} & {} &{}  &{($\%$)} &{}  &
    {($\rm{cm}^{-2} \rm{s}^{-1}$) }
     \\

    \tableline

 10 & 30 & 63 & 62.5 & 98 & 184 & 200.3 & 2 & 39 & 33.7 & 0 & 14.8 & 8.9 \\ 
 
     100 & 10 & 10 & 6.7 & 78 & 19 & 22.5 & 16 & 5 & 3.8 & 6 &  10.9 &
 6.5  \\  
     1000 & 6 & 3 & 2.7 & 59 & 8 & 7.2 & 29 & 1 & 1.4 & 12 & 6.41 & 3.8
 \\    
     10000 & 5 & 1 & 1.9 & 28 & 7 & 4.6 & 52 & 1 & 0.9 & 20 & 6.99 & 4.2
 \\ \tableline

\end{tabular}
%% Any table notes must follow the \end{tabular} command.
%%\tablenotetext{a}{Sample footnote for table~\ref{tbl-2} that}
\end{center}
\end{table}

%% If the table is more than one page long, the width of the table can vary
%% from page to page when the default \tablewidth is used, as below.  The
%% individual table widths for each page will be written to the log file; a
%% maximum tablewidth for the table can be computed from these values.
%% The \tablewidth argument can then be reset and the file reprocessed, so
%% that the table is of uniform width throughout. Try getting the widths
%% from the log file and changing the \tablewidth parameter to see how
%% adjusting this value affects table formatting.

%% The \dataset{} macro has also been applied to a few of the objects to
%% show how many observations can be tagged in a table.

\begin{table}[htbp]   
  \begin{center}
   \caption{Event summary and upmu flux limit assuming the hard
     annihilation channel is dominant. The meaning of each column is the same
    as in Table.~\ref{tab:flim_sof}. \label{tab:flim_har} }
   
   \begin{tabular}{ccccccccccccc}
  \tableline\tableline
      
        {} & {} &\multicolumn{3}{c}{Stopping} &
      \multicolumn{3}{c}{Non-showering}  &
      \multicolumn{3}{c}{Showering} &\multicolumn{2}{c}{} \\  \tableline
                      {$m_\chi$} & {$\theta$} &{$n_{\rm{obs}}$} &{$n_{\rm{BG}}$}  &
  {$F$} &{$n_{\rm{obs}}$} &{$n_{\rm{BG}}$} &{$F$} &
  {$n_{\rm{obs}}$} &{$n_{\rm{BG}}$}  &{$F$} &{$N_{90}$}  &
  {$\phi (\times10^{-15})$}

  \\
    {(GeV)} & {(deg)} &{} &{}  &
    {($\%$)} &{} &{} &{($\%$)} & {} &{}  &{($\%$)} &{}  &
    {($\rm{cm}^{-2} \rm{s}^{-1}$) }
     \\

                                       \tableline
     80.3 & 8 & 4 & 4.2 & 44 & 14 & 13.4 & 40 & 3 & 2.5 & 16 & 8.4 & 5.0  \\  
     100 & 7 & 3 & 3.3 & 42 & 10 & 9.9 & 41 & 2 & 2.2 & 17 & 6.9 & 4.1  \\ 
     1000 & 3 & 0 & 0.5 & 40 & 2 & 2.1 & 41 & 1 & 0.3 & 19 & 4.2 & 2.5 \\    
     10000 & 3 & 0 & 0.5 & 24 & 2 & 2.1 & 54 & 1 & 0.3 & 22 & 4.5 & 2.7  \\
   \tableline
       \end{tabular}
   \end{center}
  \end{table}

\end{document}